\documentclass{kluwer}    
\usepackage{graphicx}
\usepackage{lineno}
\usepackage{lscape}
\usepackage{cases}

\newcommand{\next}{\indent\newline\indent\newline}

\def\EPSFIG[#1]#2#3#4{		%
\begin{figure}[H]		%
\begin{center}			%
\includegraphics[#1]{#2}	%
\end{center}			%
\caption{#3}			%
\label{#4}			%
\end{figure}			%
}				%
\newdisplay{guess}{Conjecture}

\makeatletter
\def\@eqnnum{{\normalsize\normalcolor(\theequation)}}
\makeatother

\begin{document}

\begin{article}
\begin{opening}         
\title{Sensitivity of Displaced-Beam Scintillometer Measurements of Area-Average Heat Fluxes to Uncertainties in Topographic Heights} 
\author{Matthew \surname{Gruber}\email{matthewgruber@gi.alaska.edu}}
\author{Javier \surname{Fochesatto}\email{foch@gi.alaska.edu}}
\institute{Department of Atmospheric Science, College of Natural Sciences and Mathematics, Geophysical Institute, University of Alaska Fairbanks}
\author{Oscar \surname{Hartogensis}\email{oscar.hartogensis@wur.nl}}
\institute{Meteorology and Air Quality Group, Wageningen University, the Netherlands}

\begin{ao}
450 Cote Saint Antoine Road, Westmount, Quebec, Canada, H3Y2J9
\end{ao}


\begin{abstract}
Displaced-beam scintillometer measurements of the turbulence inner-scale length $l_o$ and refractive index structure function $C_n^2$
resolve area-average turbulent fluxes of heat and momentum through the Monin-Obukhov similarity equations. Sensitivity studies
have been produced for the use of displaced-beam scintillometers over flat terrain.
Many real field sites feature variable topography.
We develop here an analysis of the sensitivity of displaced-beam scintillometer derived sensible heat fluxes
to uncertainties in spacially distributed topographic measurements.
Sensitivity is shown to be concentrated in areas near the center of the beam 
and where the underlying topography is closest to the beam height. Uncertainty may be decreased by taking precise topographic measurements
in these areas.
\end{abstract}


\end{opening}           

\section{Introduction}

Displaced-beam scintillometers are useful to make measurements of turbulent heat fluxes since their footprint is much larger than other measurement 
devices such as sonic anemometers. Their source measurements are the index of refraction structure parameter $C_n^2$ and the 
turbulence inner scale length $l_o$, which are used to infer the turbulent
heat fluxes \cite{HILL1988L0,ANDREAS1992}.
This inference follows from equations from the Monin-Obukhov similarity hypothesis, 
which is a model for turbulent fluxes of heat and momentum in the atmospheric surface layer \cite{SORBJAN}.
Studying the sensitivity of the derived turbulent heat flux to the uncertainties in the source measurements is important, and several studies have
been published exploring the case of measurements over flat terrain \cite{ANDREAS1989,MORONI,ANDREAS1992,SOLIGNAC}.
There is debate over
whether the Monin-Obukhov similarity hypothesis can be applied over variable terrain since the canonical model is for flat terrain, however many
scintillometer field campaigns
take place over variable terrain \cite{HARTOGENSIS,MEIJNINGERPWF2002}. The turbulent fluxes are derived through equations which extend the Monin-Obukhov equations
from flat terrain to variable terrain via the application of an effective beam height \cite{HARTOGENSIS2003,KLEISSL2008,EVANS2011,GELI2012}. 
The effective beam height is calculated
assuming that the Monin-Obukhov surface layer profiles in height are valid over the slowly varying terrain. The research question to be asked
is, if one can
safely extend the Monin-Obukhov equations from flat terrain to variable terrain, then in what manner is the error propagated from the uncertainty
in topography to the derived turbulent heat flux?

Through the assumptions of the Monin-Obukhov similarity hypothesis, the sensible heat flux in the atmospheric surface layer is given by

\begin{equation}
H_S=-\rho c_p u_\star T_\star			,					\label{SENSIBLEHEAT}
\end{equation}

%

where $H_S$ is the sensible heat flux, $\rho$ is the density of air, $c_p$ is the heat capacity, $u_\star$ is the friction velocity,
and $T_\star$ is the temperature scale. Resolving $u_\star$ and $T_\star$ from $C_n^2$ and $l_o$ starts at another similarity equation
relating $\zeta\equiv z_{eff}/L$ to $T_\star$ and $u_\star$:

\begin{equation}
\zeta=\frac{\kappa G T_\star z_{eff}}{u_\star^2 T}	,	\label{zetauniversal}
\end{equation}

where $z_{eff}$ is the effective beam height of the scintillometer beam, $L$ is the 
Obukhov length, $\kappa$ is the Von K\'arm\'an constant, 
$G$ is the acceleration due to gravity, and $T$ is the temperature \cite{HARTOGENSIS2003}.
In equation \ref{zetauniversal} we omit the term due to humidity
seen in \inlinecite{ANDREAS1992} for simplicity as it has no effect on the results here.

When $\zeta>0$ we are in a stable atmospheric surface layer.
In the stable case we have

\begin{eqnarray}
\frac{C_T^2 z_{eff}^{2/3}}{T_\star^2}=a(1+c\zeta^{2/3})	\rightarrow	\pm T_\star = \sqrt{\frac{C_T^2}{a}}\frac{z_{eff}^{1/3}}{\sqrt{1+c\zeta^{2/3}}}	,	\label{CT2stable} \\
z_{eff}=\left(\int_0^1z(u)^{-2/3}G(u)du\right)^{-3/2}															,	\label{zeffstable} \\
u_\star^3=\frac{\kappa z_{eff} \epsilon}{(1+h\zeta^{3/5})^{3/2}}	\rightarrow u_\star^2=\frac{\kappa^{2/3}z_{eff}^{2/3}\epsilon^{2/3}}{(1+h\zeta^{3/5})} 	,	\label{ustarstable}
\end{eqnarray}

where $a$, $c$ and $h$ are empirical constants \cite{WYNGARD1971,WYNGARD1971phi}, 
$z(u)$ is the height profile of the beam above the ground where $u$ is the normalized distance along the beam, 
$G(u)$ is the path weighting function 
, $C_T^2$ is the temperature structure parameter which is inferred from $C_n^2$, and $z_{eff}$ is the effective beam height
assuming that the $C_T^2$ profile satisfies equation \ref{CT2stable} for any $z$ along $z(u)$ \cite{ANDREAS1992,HARTOGENSIS2003}.
The turbulent dissipation rate $\epsilon$ is directly related to $l_o$ by

\begin{eqnarray}
l_o=\frac{(9\Gamma(1/3)KD(\rho,T))^{3/4}}{\epsilon^{1/4}} ,
\end{eqnarray}

where $\Gamma$ is the Gamma function, K is the Obukhov-Corrsin constant and $D(\rho,T)$ is the thermal diffusivity of air \cite{ANDREAS1992}.

When $\zeta<0$ we are in an unstable atmospheric surface layer. 
In the unstable case we have
\begin{eqnarray}
\frac{C_T^2 z_{eff}^{2/3}}{T_\star^2}=\frac{a}{(1-b\zeta)^{2/3}}	\rightarrow	\pm T_\star = \sqrt{\frac{C_T^2}{a}}z_{eff}^{1/3}(1-b\zeta)^{1/3}	,	\label{CT2unstable} \\
z_{eff}=\frac{z_{eff}}{2b\zeta}\left(
1-
\sqrt{1-\frac{4b\zeta}{z_{eff}}\left[\int_0^1{z(u)^{-2/3}\left(1-b\zeta\frac{z(u)}{z_{eff}}\right)^{-2/3}G(u)du}\right]^{-3/2}}
\right)					,																	\label{zeffunstable} \\
u_\star^3=\frac{\kappa z_{eff} \epsilon}{(1+d(-\zeta)^{2/3})^{3/2}}	\rightarrow u_\star^2=\frac{\kappa^{2/3}z_{eff}^{2/3}\epsilon^{2/3}}{(1+d(-\zeta)^{2/3})}	,	\label{ustarunstable}
\end{eqnarray}

where $b$ and $d$ are empirical constants, and $z_{eff}$ is the effective beam height assuming that the $C_T^2$ profile
satisfies equation \ref{CT2unstable} for any $z$ along $z(u)$ \cite{HARTOGENSIS2003}.

Error is propagated from the source measurements to the derived variable $H_S$ via
the error propagation equation

\begin{equation}
 {\sigma_{f}} = \sum_{i=1}^{N}\left({\frac{\partial f}{\partial x_i}}\right)\sigma_{x_{s_i}} 
 + \sqrt{\sum_{i=1}^{N}\left({\frac{\partial f}{\partial x_i}}\right)^2{\sigma^2}_{x_{r_i}}} + \sigma_{f_{c}}	, \label{errorprop}		
\end{equation}

where the general derived variable $f$ is a function of general source measurement variables $x_{1},x_{2},...,x_{N}$ with respective systematic
error $\sigma_{x_{s_1}},\sigma_{x_{s_2}},...,\sigma_{x_{s_N}}$ and 
with respective independent Gaussian distributed uncertainties with standard deviations 
$\sigma_{x_{r_1}},\sigma_{x_{r_2}},...,\sigma_{x_{r_N}}$ as seen in \inlinecite{TAYLOR}. Computational error is given by $\sigma_{f_{c}}$.
The first and last terms represent an offset from the true solution, whereas the central term is a measure of the width of the error bars.

Error propagation can be studied with Monte-Carlo methods as in \inlinecite{MORONI}, however this is best reserved for cases where a closed-form
error propagation equation cannot be derived. We will seek a closed-form equation.
As a start, it is practical for the purpose of a sensitivity study 
to rewrite Eq. \ref{errorprop} as

\begin{equation}
 \frac{\sigma_{f}}{f} = \sum_{i=1}^{N}S_{f,x_i}\frac{\sigma_{x_{s_i}}}{x_{s_i}} 
 + \sqrt{\sum_{i=1}^{N}S_{f,x_i}^2\frac{{\sigma^2}_{x_{r_i}}}{{x_{r_i}}^2}} 	+ \frac{\sigma_{f_{c}}}{f}		,		\label{errorprop2}	
\end{equation}

where $S_{f,x}$ are unitless sensitivity functions defined by

\begin{equation}
S_{f,x} \equiv \frac{x}{f}\left(\frac{\partial f}{\partial x}\right)	.  \label{SENSITIVITYEQ}
\end{equation}

The sensitivity functions are each a measure of the portion of the error 
in the derived variable $f$ resulting from error on each individual source measurement $x$ \cite{ANDREAS1992}. Our goal is to evaluate $S_{H_S,z}(u)$, 
where the height profile $z(u)$ is distributed, hence functional derivatives will be used as in \inlinecite{GRUBERAMT}. We have

\begin{equation}
S_{H_S,z}(u)=\frac{z(u)}{H_S}\left(\frac{\delta H_S}{\delta z}(u)\right)	.\label{SHszu}
\end{equation}

The sensitivity function $S_{H_S,z}$ for this measurement strategy over flat terrain has been evaluated in \inlinecite{GRUBERBLM}.

We will solve for the sensitivity function in equation \ref{SHszu}
for stable conditions in section 2. In section 3 we will solve for the sensitivity function in equation \ref{SHszu} for 
unstable conditions. In section 4 we will
apply the sensitivity function to the topography of a real field site. We conclude in section 5.

\section{Stable case ($\zeta>0$)}

Here we can combine equations \ref{zetauniversal}, \ref{CT2stable} and \ref{ustarstable} to arrive at

\begin{eqnarray}
\zeta^2+c\zeta^{8/3}=\hat{A}(1+h\zeta^{3/5})^2z_{eff}^{4/3}	,	\label{zetastable}
\end{eqnarray}

where 

\begin{eqnarray}
 \hat{A}\equiv\frac{\kappa^{2/3}g^2 C_T^2}{T^2a\epsilon^{4/3}}	.	\label{Ahatstable}
\end{eqnarray}

\begin{landscape}

Since $z_{eff}$ is determined independently from equation \ref{zeffstable}, equation \ref{zetastable} is a single equation in the 
single unknown $\zeta$, where all other variables in the equation are measured. From this equation we can determine the 
variable inter-dependency as illustrated in the tree diagram seen in figure \ref{TREE1}.

\begin{figure}
  \centering
  \includegraphics[width=18cm]{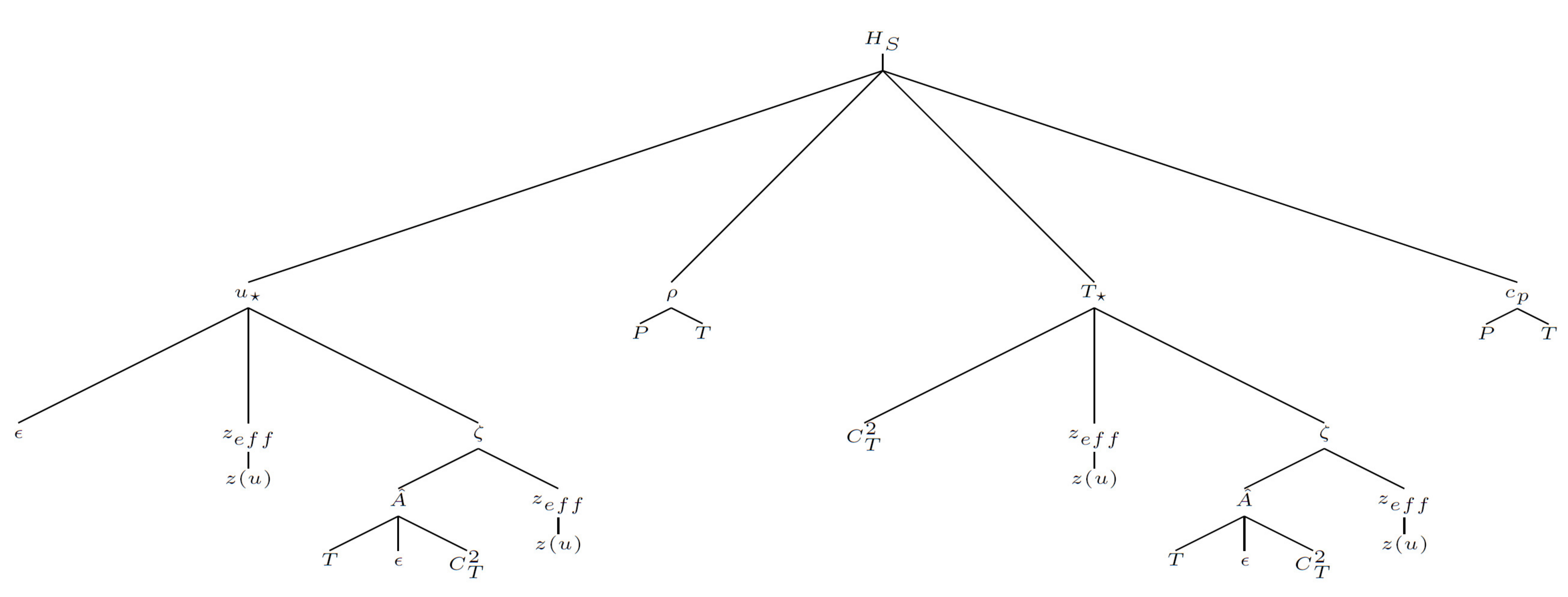}
  \caption[Variable inter-dependency tree diagram.]
    {Variable inter-dependency tree diagram for $H_S$ under stable conditions corresponding to $\zeta>0$. The tree
    diagram for the dependence of $\epsilon$ on $P$, $T$ and $l_o$ is omitted.}
  \label{TREE1}
\end{figure}

From the tree diagram seen in figure \ref{TREE1}, we have

\small
\begin{eqnarray}
\left(\frac{\delta H_S}{\delta z}(u)\right)=\frac{\partial H_S}{\partial T_\star}
\left[\left(\frac{\partial T_\star}{\partial z_{eff}}\right)_\zeta+\left(\frac{\partial T_\star}{\partial \zeta}\right)\left(\frac{\partial \zeta}{\partial z_{eff}}\right)\right]\left(\frac{\delta z_{eff}}{\delta z}(u)\right)+
\frac{\partial H_S}{\partial u_\star}
\left[\left(\frac{\partial u_\star}{\partial z_{eff}}\right)_\zeta+\left(\frac{\partial u_\star}{\partial \zeta}\right)\left(\frac{\partial \zeta}{\partial z_{eff}}\right)\right]\left(\frac{\delta z_{eff}}{\delta z}(u)\right) .
\end{eqnarray}
\normalsize

Many of these derivatives follow directly from the definitions. For $\left(\frac{\partial \zeta}{\partial z_{eff}}\right)$
we must implicitly differentiate both sides of equation \ref{zetastable} to arrive at

\begin{eqnarray}
\left(\frac{\partial \zeta}{\partial z_{eff}}\right)=
\frac{4(\zeta^2+c\zeta^{8/3})}
{3z_{eff}\left(2\zeta+\frac{8}{3}c\zeta^{5/3}-\frac{6}{5}h\left(\frac{\zeta^{8/5}-c\zeta^{34/15}}{1+h\zeta^{3/5}}\right)\right)} .
\end{eqnarray}

We then achieve

\small
\begin{eqnarray}
\frac{z(u)}{H_S}\left(\frac{\delta H_S}{\delta z}(u)\right)=
\left[\frac{2}{3}-\left(\frac{c\zeta^{2/3}}{3(1+c\zeta^{2/3})}+\frac{3h\zeta^{3/5}}{10(1+h\zeta^{3/5})}\right)
\left(\frac{\frac{4}{3}(1+c\zeta^{2/3})}
{2+\frac{8}{3}c\zeta^{2/3}-\frac{6}{5}h\left(\frac{\zeta^{3/5}-c\zeta^{19/15}}{1+h\zeta^{3/5}}\right)}\right)
\right]
\frac{z(u)^{-2/3}G(u)}{\int_0^1 z(u)^{-2/3}G(u)du}	.	\label{SHzstable}
\end{eqnarray}
\normalsize

\section{Unstable case ($\zeta<0$)}

 Here we combine equations \ref{zetauniversal}, 
\ref{CT2unstable} and \ref{ustarunstable}; we achieve

\begin{eqnarray}
\frac{\zeta}{z_{eff}}=-\breve{A}\Phi(\zeta)	,	\label{zetaunstable}
\end{eqnarray}

\end{landscape}

where 

\begin{eqnarray}
\breve{A}=\frac{\sqrt{\kappa}g^{3/2}(C_T^2)^{3/4}}{T^{3/2}\epsilon a^{3/4}}	,	\label{Ahatunstable} \\
\Phi(\zeta)=\sqrt{\frac{(1-b\zeta)(1+d(-\zeta)^{2/3})^3}{(-\zeta)}}		.	\label{Phi}
\end{eqnarray}

In the unstable case, $z_{eff}$ is coupled to $\zeta$ through equation \ref{zeffunstable}. We input equation
\ref{zetaunstable} into equation \ref{zeffunstable} to arrive at

\footnotesize
\begin{eqnarray}
\zeta=\frac{1}{2b}
\left(
1-
\sqrt{1+4b\breve{A}\Phi(\zeta)\left[\int_0^1{\left(z(u)+bz(u)^2\breve{A}\Phi(\zeta)\right)^{-2/3}G(u)du}\right]^{-3/2}}
\right)									.							\label{zetaunstablefinal}
\end{eqnarray}
\normalsize

This is a single equation in the single unknown $\zeta$, where all other variables are measured.
This equation is in fixed point form and can be solved numerically via fixed point recursion as seen in \inlinecite{ITERATIVE} and in
\inlinecite{FIXEDPOINT}. The variable interdependency is mapped out
in the tree diagram seen in figure \ref{TREE2}.

\begin{figure}[H]
  \centering
  \includegraphics[width=8cm]{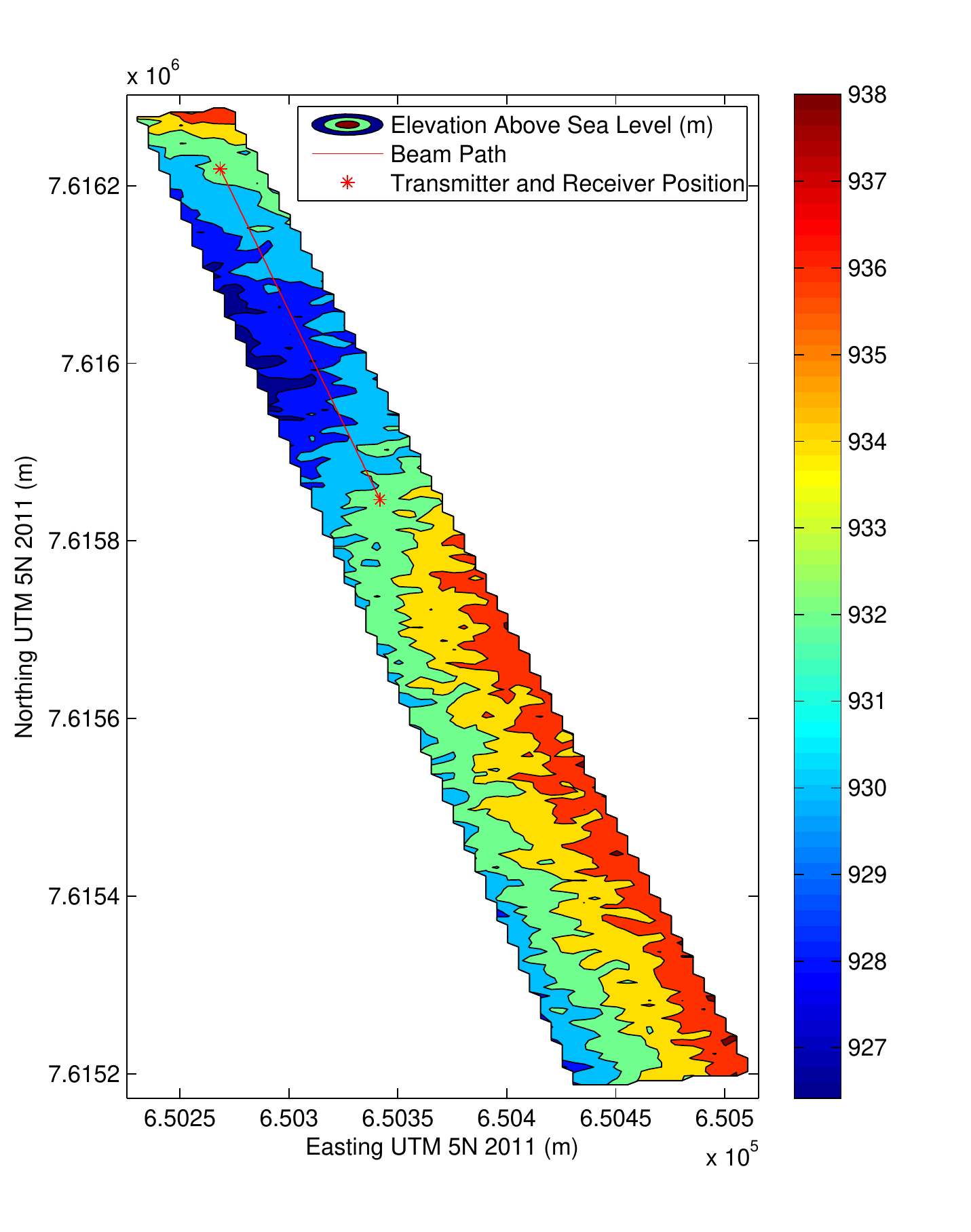}
  \caption[Imnavait basin.]
    {Imnavait basin topography and beam path. Imnavait basin is located on the North Slope of Alaska. The beam emitter and receiver
    are raised on tripods $1.8m$ high.}
  \label{IMNAVIATPATH}
\end{figure}

\begin{landscape}

\begin{figure}
  \centering
  \includegraphics[width=20cm]{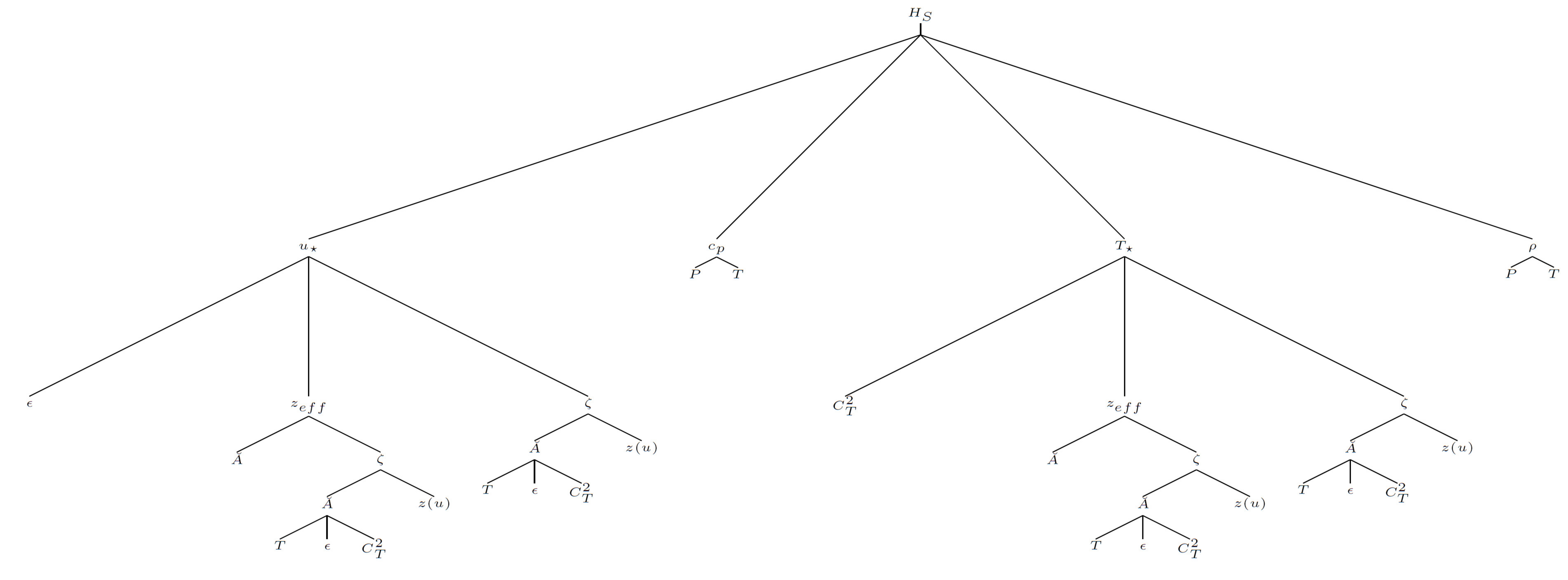}
  \caption[Variable inter-dependency tree diagram.]
  {Variable inter-dependency tree diagram for $H_S$ under unstable conditions corresponding to $\zeta<0$.
  The tree diagram for the dependence of $\epsilon$ on $P$, $T$ and $l_o$ is omitted.}
  \label{TREE2}
\end{figure}

From the tree diagram seen in figure \ref{TREE2} we have

\small
\begin{eqnarray}
\left(\frac{\delta H_S}{\delta z}(u)\right)=\frac{\partial H_S}{\partial T_\star}
\left[\left(\frac{\partial T_\star}{\partial \zeta}\right)_{z_{eff}}+\left(\frac{\partial T_\star}{\partial z_{eff}}\right)\left(\frac{\partial z_{eff}}{\partial \zeta}\right)\right]   \left(\frac{\delta \zeta}{\delta z}(u)\right)+
\frac{\partial H_S}{\partial u_\star}
\left[\left(\frac{\partial u_\star}{\partial \zeta}\right)_{z_{eff}}+\left(\frac{\partial u_\star}{\partial z_{eff}}\right)\left(\frac{\partial z_{eff}}{\partial \zeta}\right)\right]   \left(\frac{\delta \zeta}{\delta z}(u)\right) .
\label{UNSTABLEPARTIALS}
\end{eqnarray}
\normalsize

From equation \ref{zetaunstable} we have

\begin{eqnarray}
\frac{\partial z_{eff}}{\partial \zeta}=z_{eff}\left(\frac{1}{\zeta}-\frac{\left(\frac{\partial \Phi}{\partial \zeta}\right)}{\Phi(\zeta)}\right) ,
\end{eqnarray}

and from equation \ref{Phi} we have

\begin{eqnarray}
\left(\frac{\partial \Phi}{\partial \zeta}\right)=\frac{1}{2\Phi(\zeta)}
\left[\frac{(1+d(-\zeta)^{2/3})^3+2d\zeta(1-b\zeta)(1+d(-\zeta)^{2/3})^2(-\zeta)^{-1/3}}
{\zeta^2}\right] .
\end{eqnarray}

From equation \ref{zetaunstablefinal} we have

\begin{eqnarray}
\left(\frac{\delta \zeta}{\delta z}(u)\right)=
\frac{-\left(\frac{\delta f}{\delta z}(u)\right)_\zeta}
{\left(\frac{\partial f}{\partial \zeta}\right)+4b(1-2b\zeta)}	,	\label{deltazetadeltazunstable1}
\end{eqnarray}

where

\begin{eqnarray}
f\left(\breve{A},z(u),\zeta(\breve{A},z(u))\right)\equiv
1+4b\breve{A}\Phi(\zeta)\left[\int_0^1{\left(z(u)+b\breve{A}\Phi(\zeta)z(u)^2\right)^{-2/3}G(u)du}\right]^{-3/2} .
\end{eqnarray}

We thus have

\begin{eqnarray}
\frac{z(u)}{H_S}\left(\frac{\delta H_S}{\delta z}(u)\right)=
z(u) \left[\frac{2}{3}\left(\frac{1}{\zeta}-\frac{\left(\frac{\partial \Phi}{\partial \zeta}\right)}{\Phi(\zeta)}\right)+
\frac{d}{3(1+d(-\zeta)^{2/3})(-\zeta)^{1/3}}
-\frac{b}{3(1-b\zeta)}
\right]
\left(\frac{\delta \zeta}{\delta z}(u)\right) 		,	\label{SHzunstable}
\end{eqnarray}

from equation \ref{UNSTABLEPARTIALS}, where $\left(\frac{\delta \zeta}{\delta z}(u)\right)$ is solved through equation \ref{deltazetadeltazunstable1} as

\tiny
\begin{eqnarray}
\left(\frac{\delta \zeta}{\delta z}(u)\right)=
\frac{-\breve{A}\Phi(\zeta)\left(z(u)+b\breve{A}\Phi(\zeta)z(u)^2\right)^{-5/3}\left(1+2b\breve{A}z(u)\Phi(\zeta)\right)G(u)}
{\breve{A}\left(\frac{\partial \Phi}{\partial \zeta}\right)\left[\int_0^1{\left(z(u)+b\breve{A}\Phi(\zeta)z(u)^2\right)^{-2/3}G(u)du}\right]+
b\breve{A}^2\Phi(\zeta)\left(\frac{\partial \Phi}{\partial \zeta}\right)
\left[\int_0^1{\left(z(u)+b\breve{A}\Phi(\zeta)z(u)^2\right)^{-5/3}z(u)^2G(u)du}\right]
+(1-2b\zeta)\left[\int_0^1{\left(z(u)+b\breve{A}\Phi(\zeta)z(u)^2\right)^{-2/3}G(u)du}\right]^{5/2}	.	\label{deltazetadeltazunstable2}
}
\end{eqnarray}
\normalsize

\section{Results}

We have solved the sensitivity function $S_{{H_S},z}(u)$ which depends on $z(u)$ and $\zeta$ for both
stable conditions where $\zeta>0$ and for unstable conditions where $\zeta<0$. 
In order to visualize the sensitivity function, we will demonstrate it with a field site beam path height profile $z(u)$. 
We use the Imnavait basin field site seen in figure \ref{IMNAVIATPATH}.

\end{landscape}

\begin{figure}[H]
  \centering
  \includegraphics[width=12cm]{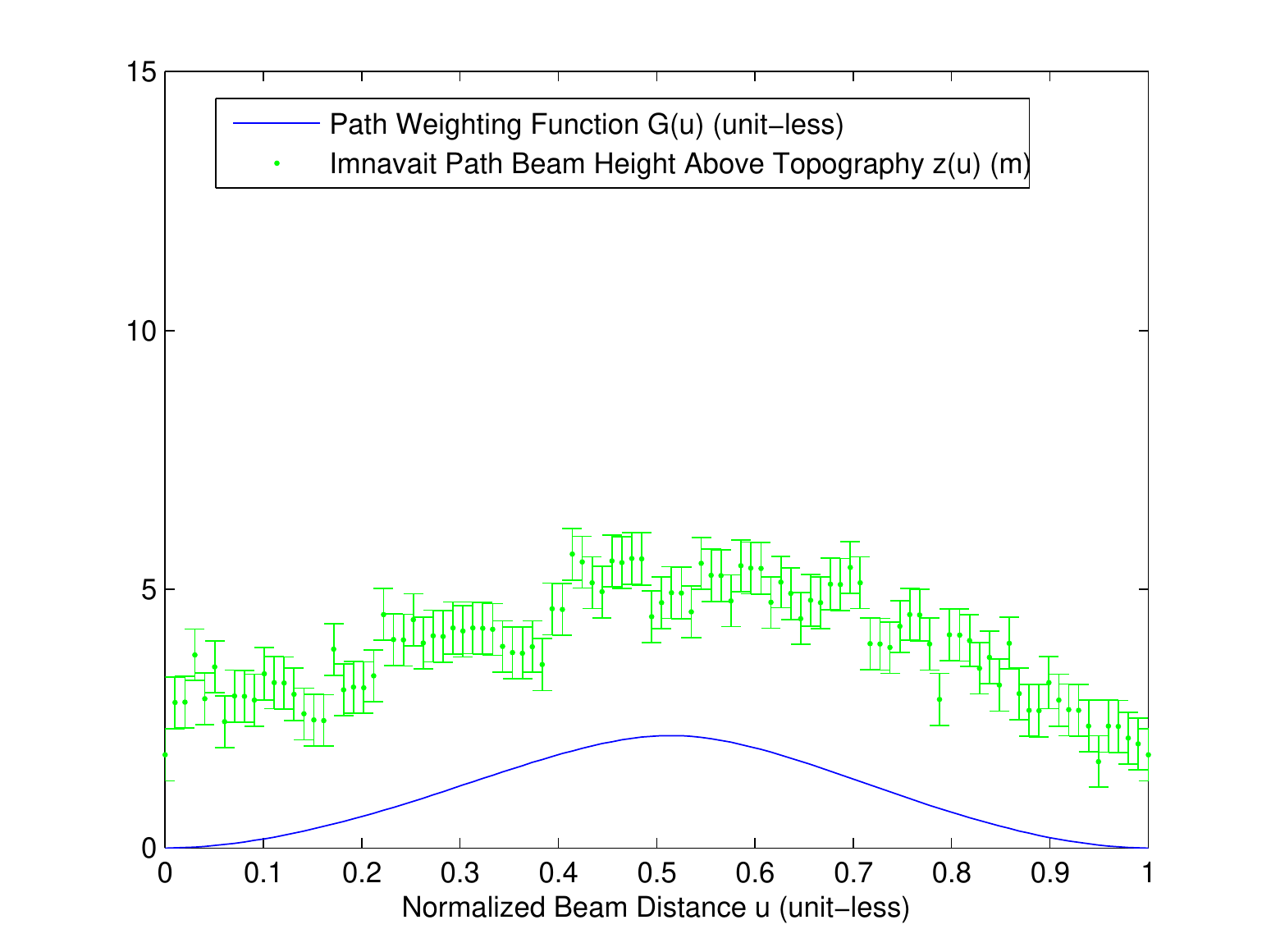}
  \caption[Imnavait basin.]
    {Imnavait basin beam path heights above topography. Imnavait basin is located on the North Slope of Alaska. The beam emitter and receiver
    are raised on tripods $1.8m$ high.}
  \label{zG}
\end{figure}

\begin{figure}[H]
  \centering
  \includegraphics[width=10cm]{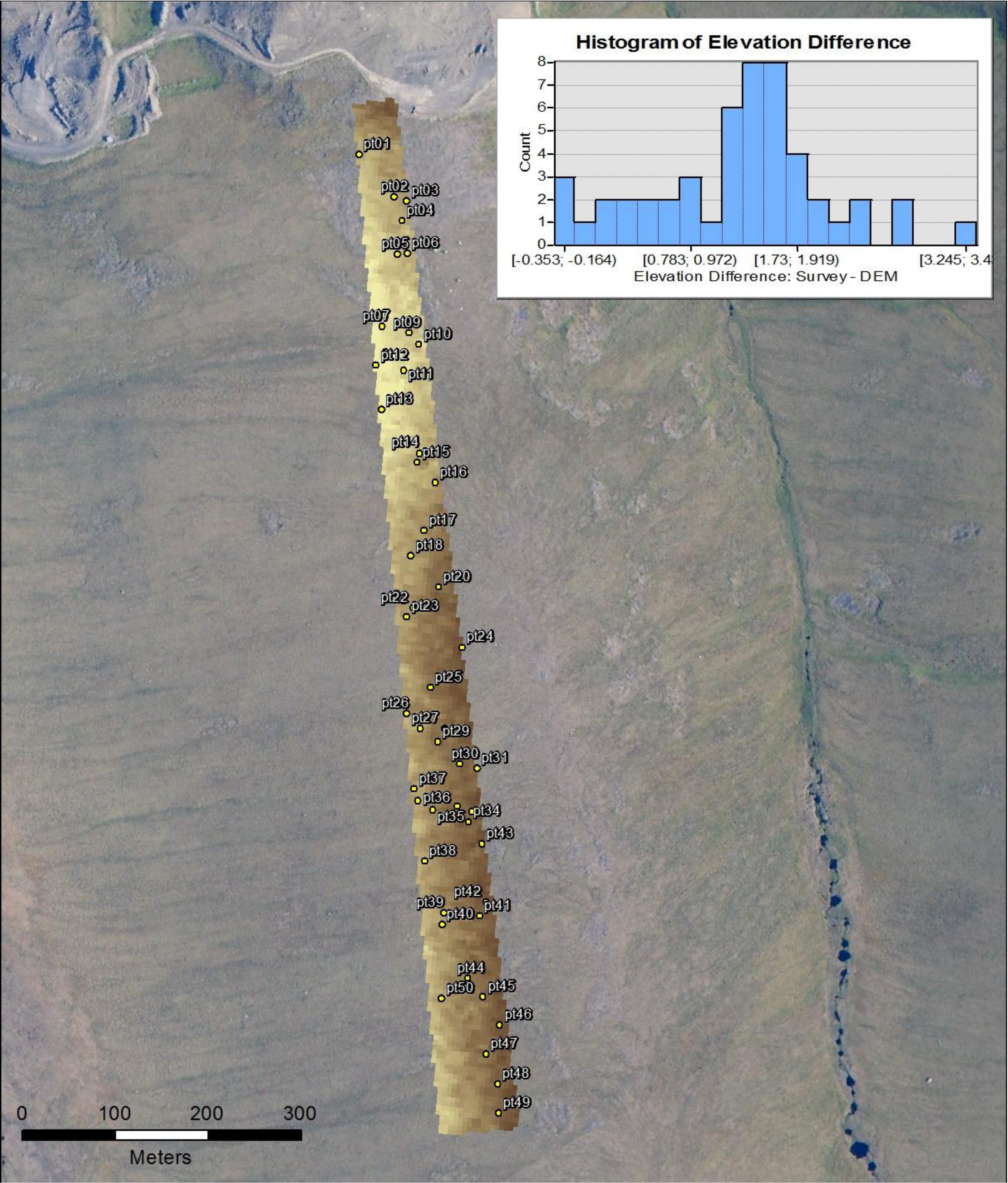}
  \caption[Imnavait basin.]
    {Imnavait basin ground truth GPS measurements. Overlaid is a histogram of the difference between the GPS measurements
    and the digital elevation map used in figures \ref{IMNAVIATPATH} and \ref{zG}. The standard deviation is approximately $0.5m$.}
  \label{GPS}
\end{figure}

This field site has a beam height $z(u)$ as seen in figure \ref{zG}. A random error component of $0.5m$ is used in figure \ref{zG} since the difference between ground truth GPS measurements and the digital elevation map used
has a standard deviation of approximately $0.5m$ as seen in figure \ref{GPS}. Note that the magnitude of the error does not influence the results of this study.

\begin{figure}[H]
  \centering
  \includegraphics[width=15cm]{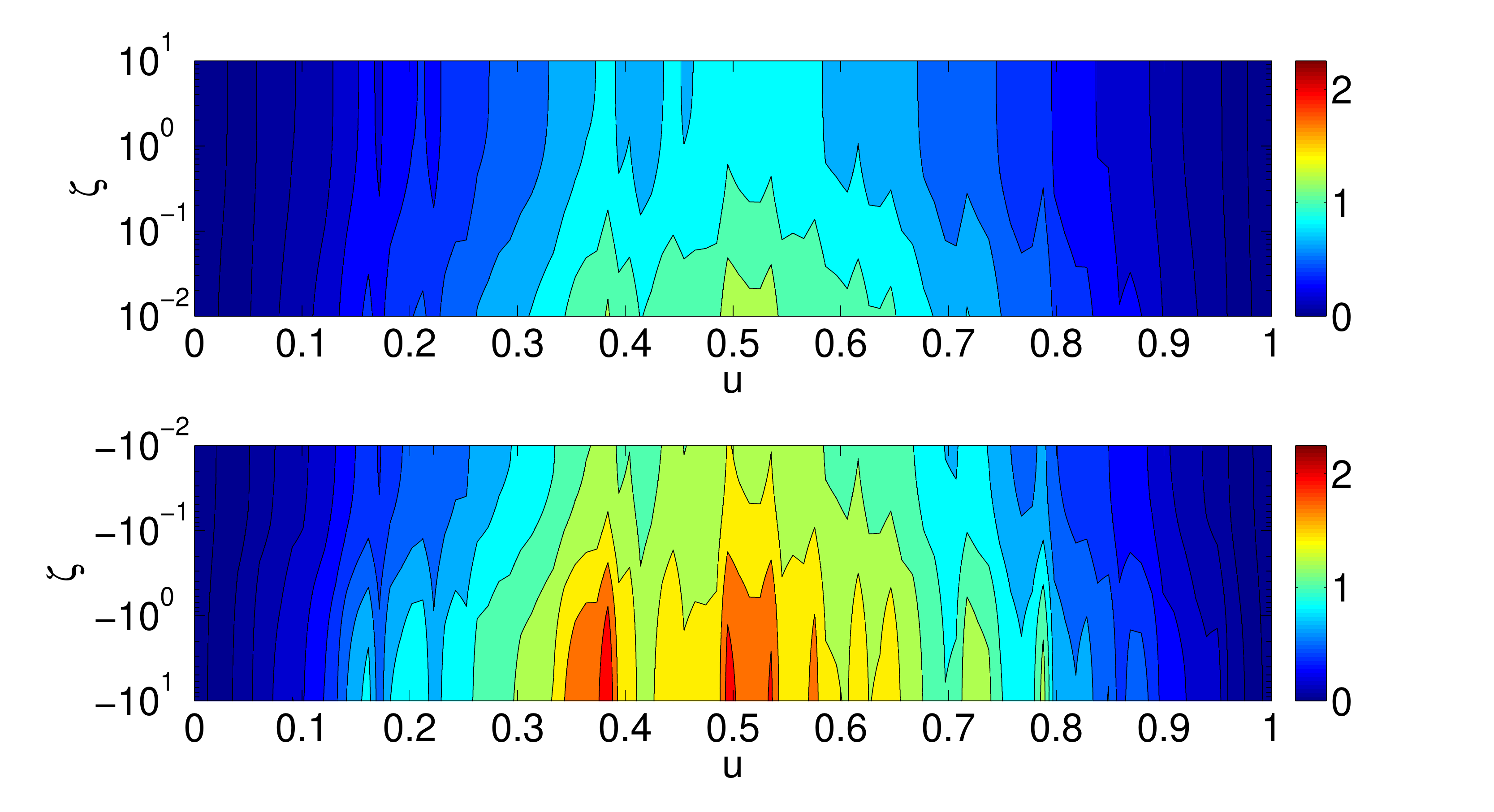}
  \caption[Variable inter-dependency tree diagram.]
    {Sensitivity function $S_{{H_S},z}(u)$ over the Imnavait basin field site seen in figure \ref{IMNAVIATPATH} and in figure \ref{zG}.}
  \label{SHsz}
\end{figure}

For the Imnavait basin path $z(u)$ seen in figure \ref{zG}, 
taking equation \ref{SHzstable} for the stable case, and equations \ref{SHzunstable} and \ref{deltazetadeltazunstable2} for the unstable case,
we arrive at the sensitivity function seen in figure \ref{SHsz}. Note that values of $\breve{A}$ for a given $\zeta$ in equation \ref{deltazetadeltazunstable2}
follow from the numerical solution to equation \ref{zetaunstablefinal}.

The sensitivity gets higher from stable conditions to unstable conditions, and it is focused in areas near the center of the beam path, 
as well as in areas of topographic protrusion. Note the clear spikes in sensitivity at $u=0.38$ and at $u=0.5$ in figure \ref{SHsz}.
These points in the path correspond to local minima in $z(u)$ as seen in figure \ref{zG}. This makes sense since $C_T^2$ decreases
nonlinearly in height above the ground, most rapidly near the surface. In areas where the beam approaches the ground, the gradient
in $C_T^2$ is higher as seen in equations \ref{CT2stable} and \ref{CT2unstable}, therefore uncertainties in the actual height of the beam
in those areas will translate into high uncertainties in the derived variables.

\begin{figure}[H]
  \centering
  \includegraphics[width=15cm]{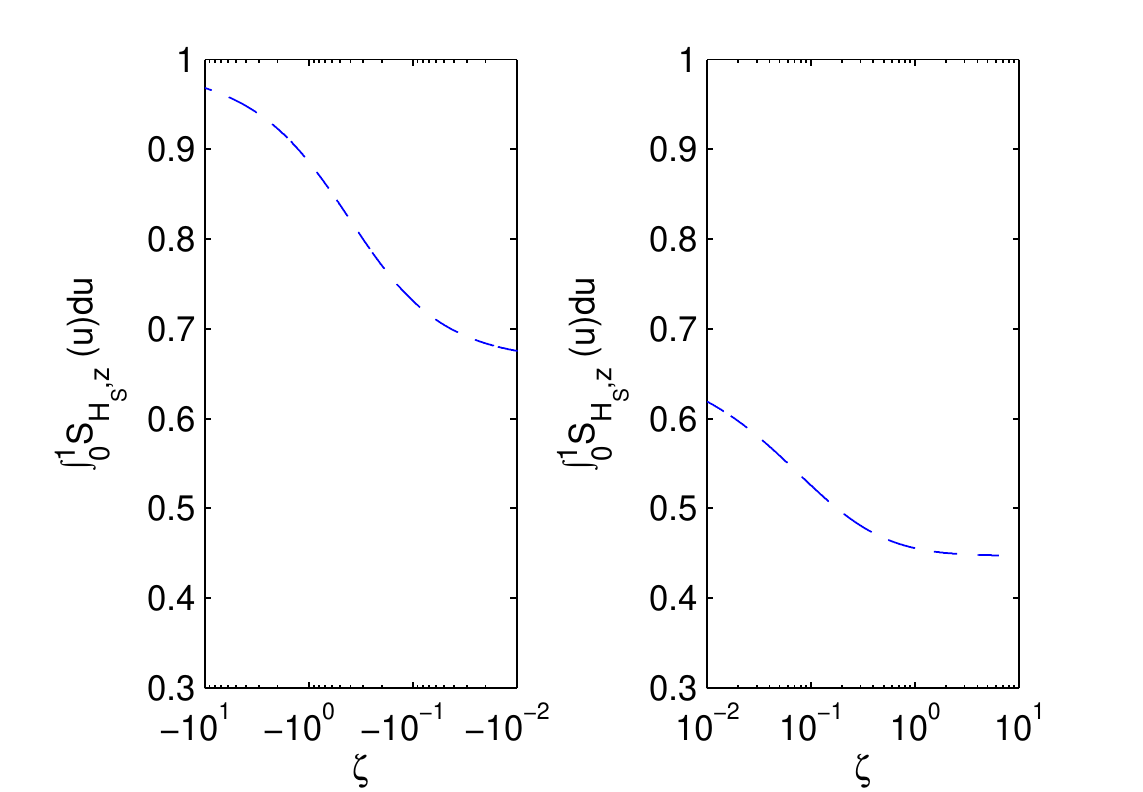}
  \caption[Variable inter-dependency tree diagram.]
    {Sensitivity function $S_{{H_S},z}(u)$ integrated over $u$ for any field site.}
  \label{SHszint}
\end{figure}

Note that if we consider a constant ratio of $\frac{\sigma_{z}(u)}{z(u)}$, the term in, for example Eq. (\ref{errorprop2}), can be re-written as

\begin{equation}
\int\limits_0^1{\frac{\sigma_{z}(u)}{z(u)}S_{H_S,z}(u)du}
= \frac{\sigma_{z}(u)}{z(u)}\left[\int\limits_0^1{S_{H_S,z}(u)du}\right]  ,  \label{CONSTRATIO}
\end{equation}

where the term in square brackets is plotted in figure \ref{SHszint}.

This result is the average value of the sensitivity along the whole length of the path, and it converges to the same result for all paths.
The sensitivity function in figure \ref{SHszint} is compatible with the one-dimensional  sensitivity function for flat terrain
seen in \inlinecite{GRUBERBLM}.

\section{Conclusion}

Using the effective beam height extension of the Monin-Obukhov similarity equations to variable terrain, we have solved for how
uncertainty in topographic heights propagates to displaced-beam scintillometer measurements of turbulent heat fluxes over any field site. We have found
that uncertainty is concentrated in areas around topographic protuberances, as well as near the center of the beam path
as seen in figures \ref{zG} and \ref{SHsz}. The local
sensitivity can easily approach values of $200\%$ in unstable conditions as seen in figure \ref{SHsz}, but the average sensitivity over the beam path never 
exceeds $100\%$ as seen in figure \ref{SHszint}. These results carry important ramifications in the selection of beam paths, in the calculation of uncertainty, and in
the type of topographic data used. It may be that for many scintillometer beam paths, in order to achieve reasonable uncertainty we must
use high precision LIDAR topographic data in order to reduce what is likely the greatest contributor to overall uncertainty.

It is interesting that the average value of the sensitivity function 
$S_{H_S,z}(u)$ over the beam path reduces to the identical sensitivity function $S_{H_S,z}$
for diplaced-beam scintillometers over flat terrain as seen in figure \ref{SHszint} and in \inlinecite{GRUBERBLM}. We have essentially
expanded the flat terrain sensitivity function first explored in \inlinecite{ANDREAS1992}
from one dimension to two, and then we averaged through one dimension
to arrive back at the original one-dimensional sensitivity function. Future work should perhaps focus on applying this type of sensitivity
analysis on large aperture scintillometers
using the Businger-Dyer relation to obtain path averaged $u_\star$, $T_\star$ and $H_S$ measurements. Additional work should be performed
for scintillometer paths which are below the blending height over heterogeneous terrain 
\cite{WIERINGA1986,MASON1987,CLAUSSEN1990,CLAUSSEN1995,MEIJNINGERPWF2002,HARTOGENSIS2003,LILU}.

\acknowledgements
Matthew Gruber thanks the Geophysical Institute for its support during his Master's degree program in Atmospheric Sciences at the
University of Alaska Fairbanks. We thank 
Jason Stuckey and  Randy Fulweber at ToolikGIS, Chad Diesinger at Toolik Research Station, and Matt Nolan at the Institute for Northern Engineering
for the digital elevation map of Imnavait, data support, field site GPS measurements and figure \ref{GPS}.

%
%




\end{article}
\end{document}